\documentclass[preprint]{aastex}
\usepackage{graphicx}

\usepackage{caption}
\usepackage{amsmath, amssymb}
\usepackage{calligra}
\usepackage{url}
\usepackage{natbib}
\usepackage{color}
\usepackage{pdflscape}
\usepackage{hyperref}
\usepackage{tikz} 
\usepackage{subfigure}
\usepackage{longtable}
\usepackage{array}
\usepackage{float}
\usetikzlibrary{shapes,arrows,fit,backgrounds}
\let\new=\newcommand
\new{\diff}{{\rm d}}

\slugcomment{First Belgo-Indian Network for Astronomy \& astrophysics (BINA) workshop, Nainital (India), 15-18 November 2016}

\shorttitle{Physics and observations of TDEs}
\shortauthors{Mangalam \& Mageshwaran}

\begin{document}

\title{Physics and observations of tidal disruption events}

\author{A.~Mangalam$^{\dagger}$,~~~~T.~Mageshwaran$^{\ddagger}$}

\affil{Indian Institute of Astrophysics, Bangalore-560034, India}

\email{ $^{\dagger}$mangalam@iiap.res.in, $^{\ddagger}$mageshwaran@iiap.res.in}

\begin{abstract}

We describe a model of tidal disruption events (TDEs) with input physical parameters that include the black hole (BH) mass $M_{\bullet}$, the specific orbital energy $E$, the angular momentum $J$, the star mass $M_{\star}$ and radius $R_{\star}$. We calculate the rise time of the TDEs, the peak bolometric luminosity in terms of these physical parameters and a typical light curve of TDEs for various All Sky Survey (ASS) and Deep Sky Survey (DSS) missions. We then derive the expected detection rates and discuss the follow up of TDEs through observations in various spectral bands from X-rays to radio wavelengths.  

\end{abstract}

\section{Introduction}

If a star passes within the tidal radius $\displaystyle{r_{t}\sim R_{\star} (M_{\bullet}/ M_{\star})^{1/3}}$ of the galaxy's central BH, the BH's tidal gravity exceeds the star's self-gravity and it is tidally disrupted \citep{1988Natur.333..523R}. The energy of the disrupted debris $E_d$ depends on the pericenter of the star orbit $r_{p}(E,\hspace{1mm}J,\hspace{1mm}M_{\bullet},\hspace{1mm} M_{\star})$ (\citealt{2015ApJ...814..141M}; hereafter MM15). The stars are tidally captured if the angular momentum, $J\leq J_{lc}(E,r_{t})$ where $J_{lc}(E,r_{t})=\sqrt{2r^2_t(\Phi (r_t)-E)}$ is the loss cone angular momentum  \citep{1976MNRAS.176..633F}, the maximum value of $J$ is $J_{lc}(E,r_{t})$. As $J_{lc}(E,r_{t})\geq 0$, the maximum value of energy is $E_{m}= \Phi (r_t)$. We define the dimensionless energy $\bar{e}=E/E_m$ and angular momentum $\ell=J/J_{lc}$ and the constraints in the energy and angular momentum phase, $\bar{e}_h=r_t/r_h< \bar{e}<1$ and $0<\ell <1$ where $r_h=GM_{\bullet}/\sigma^2$ and $\sigma$ is the stellar velocity dispersion; in the following sections we use the scaling relation $M_{\bullet}-\sigma$ relation given by \citet{2005SSRv..116..523F} 

\begin{equation}
M_{\bullet}=1.66\times 10^8 M_{\odot}\left(\frac{\sigma}{200~{\rm Km ~ sec^{-1}}}\right)^{4.86}.
\end{equation}

The fractional radius upto which the debris is bound to the BH at the moment of disruption is
\begin{equation}
x_{\ell}={\rm Min[}1,~r_p^2E/(2kGM_{\bullet}R_{\star})\rm{]}, 
\end{equation}
where $k$ is spin up factor taken to be 3. In terms of dimensionless variables, $x_{\ell}$ is given by

\begin{equation}
x_{\ell}={\rm Min}\left[1,~\left(\frac{M_{\bullet}}{M_{\star}}\right)^{\frac{1}{3}}\frac{\bar{e}}{2k}\left\{\frac{2\ell^2 (1-\bar{e})}{1+\sqrt{1-4\ell^2\bar{e}(1-\bar{e})}}\right\}^2 \right].
\end{equation}

The mass in-fall rate of debris depends on the internal structure and properties of the star, and follows the $t^{-5/3}$ law at the late stages of its evolution. The mass fallback rate given in MM15 is approximated by

\begin{equation}
\dot{M}_{fb}=\frac{2}{3}\frac{M_{\star}}{t_m}(1+x_{\ell}) (-0.0175~ x^2+0.0175~ {\rm Exp}[-4.216 (x^2-1)])\tau^{-\frac{5}{3}},
\end{equation}

where $\tau=t/t_m$, $x=x_{\ell}-\tau^{-2/3}(1+x_{\ell})$ and $t_m$ is time period of inner bound debris given by

\begin{equation}
t_m=0.041~ {\rm days}~~m^{0.7}\bar{e}^{-\frac{3}{2}}\left[1+\frac{1}{x_{\ell}}\right]^{-\frac{3}{2}},
\end{equation}

where $m=M_{\star}/M_{\odot}$ and $R_{\star}=R_{\odot}m^{0.8}$ is considered \citep{1994sse..book.....K}. The peak mass fallback rate and the corresponding rise time given in MM15 are approximated by 

\begin{equation}
\dot{M}_p=5.6 \times 10^{29} {\rm g~sec^{-1}}~~\left[ m^{0.3}\bar{e}^{\frac{3}{2}} \left[1+\frac{1}{x_{\ell}}\right]^{\frac{3}{2}}(0.05+0.16x_{\ell}+0.197 x_{\ell}^2-0.026 x_{\ell}^3)\right],
\label{mpeak}
\end{equation}

\begin{equation}
t_p=0.041~ {\rm days}~~\left[m^{0.7}\bar{e}^{-\frac{3}{2}}\left[1+\frac{1}{x_{\ell}}\right]^{-\frac{3}{2}}(2.66-0.625x_{\ell}-0.177 x_{\ell}^2+0.164 x_{\ell}^3)\right].
\label{tpeak}
\end{equation}

\section{Accretion disk luminosity}
\label{lum}

 In this section, we consider a steady accretion model with $\alpha$ viscosity and without fallback. The sub-Eddington disk with $\alpha$ viscosity is a thin disk whose temperature profile, bolometric luminosity and luminosity in spectral band $\{\nu_{l},~\nu_{h}\}$ assuming a blackbody emission, are given by (MM15)

\begin{equation}
\sigma_{SB}T_e^4=\frac{3}{8}\frac{GM_{\bullet}\dot{M}}{r^3}\left(1-\sqrt{\frac{r_{ISCO}}{r}}\right),
\end{equation}

\begin{equation}
L_b=\frac{3\pi}{4}\frac{GM_{\bullet}\dot{M}}{r_{ISCO}}\left[1-\frac{r_{ISCO}}{r_{out}}-\frac{2}{3}\left(1-\left(\frac{r_{ISCO}}{r_{out}}\right)^{\frac{3}{2}}\right)\right],
\label{bolum}
\end{equation} 

and

\begin{equation}
L_{\{\nu_{l},~\nu_h\}}(T)=\int_{r_{in}}^{r_{out}} {\rm dr} 2\pi r \int_{\nu_{l}(1+z)}^{\nu_{h}(1+z)} {\rm d\nu}~~ \frac{2h}{c^2} \frac{\nu^3}{{\rm Exp}[\frac{h\nu}{k_B T}]-1},
\label{speclum}
\end{equation}

where $h$ is the Planck constant, $c$ the speed of light, $k_{B}$ the Boltzmann constant, $\sigma_{SB}$ the Stefan-Boltzmann constant, $z$ is the redshift, $r_{ISCO}=(G M_{\bullet}/c^2)Z(j)$ is the inner radius with $Z(j)$ given in \citet{1972ApJ...178..347B}, where $j$ is the black hole spin and the outer radius $r_{out}$ is taken to be the  circularization radius $r_c=r_tf(\bar{e},\ell)$ with

\begin{equation}
f(\bar{e},~\ell)=\frac{4 \ell^2 (1-\bar{e})}{1+\sqrt{1-4\ell^2 \bar{e} (1-\bar{e})}}\left[1-\frac{2 \ell ^2 \bar{e}(1-\bar{e})}{1+\sqrt{1-4\ell^2 \bar{e} (1-\bar{e})}}\right].
\label{fel}
\end{equation}

Similar to \citet{2009MNRAS.400.2070S}, we assume that the accretion rate $\dot{M}$ is similar to the fallback rate which implies that the peak bolometric luminosity using eqn (\ref{mpeak}) is calculated to be

\begin{multline}
L_b^{peak}=5.1 \times 10^{49} {\rm erg~sec^{-1}} \frac{m^{0.3}\bar{e}^{\frac{3}{2}}}{Z(j)}\left[1+\frac{1}{x_{\ell}}\right]^{\frac{3}{2}}(0.05+0.16x_{\ell}+0.197 x_{\ell}^2-0.026 x_{\ell}^3)\\ \left[1-\frac{r_{ISCO}}{r_{out}}-\frac{2}{3}\left(1-\left(\frac{r_{ISCO}}{r_{out}}\right)^{\frac{3}{2}}\right)\right].
\label{lpeak}
\end{multline}

\begin{figure}[ht]
\centering
\subfigure[]{\includegraphics[width=8cm]{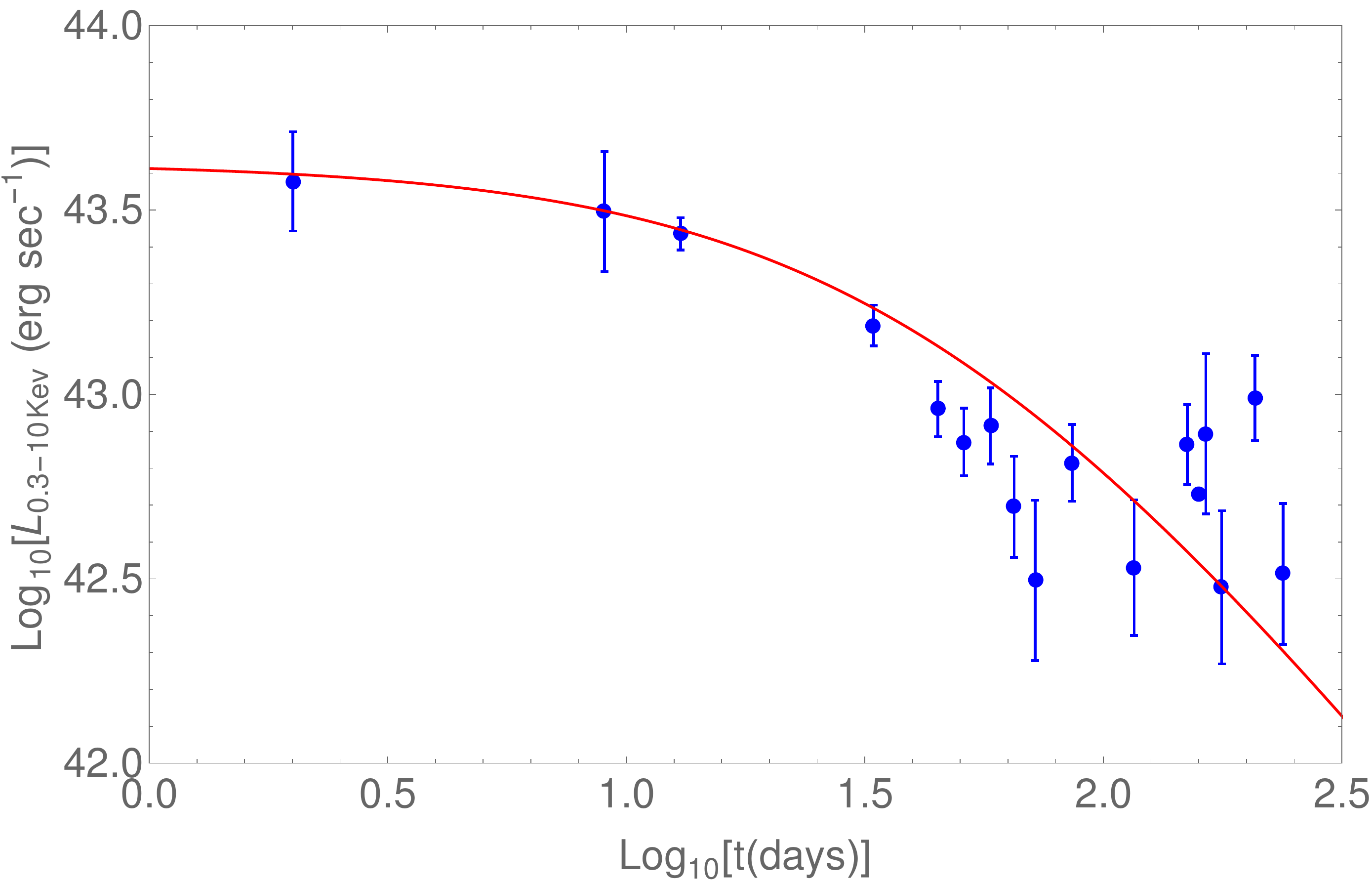}}
\subfigure[]{\includegraphics[width=8cm]{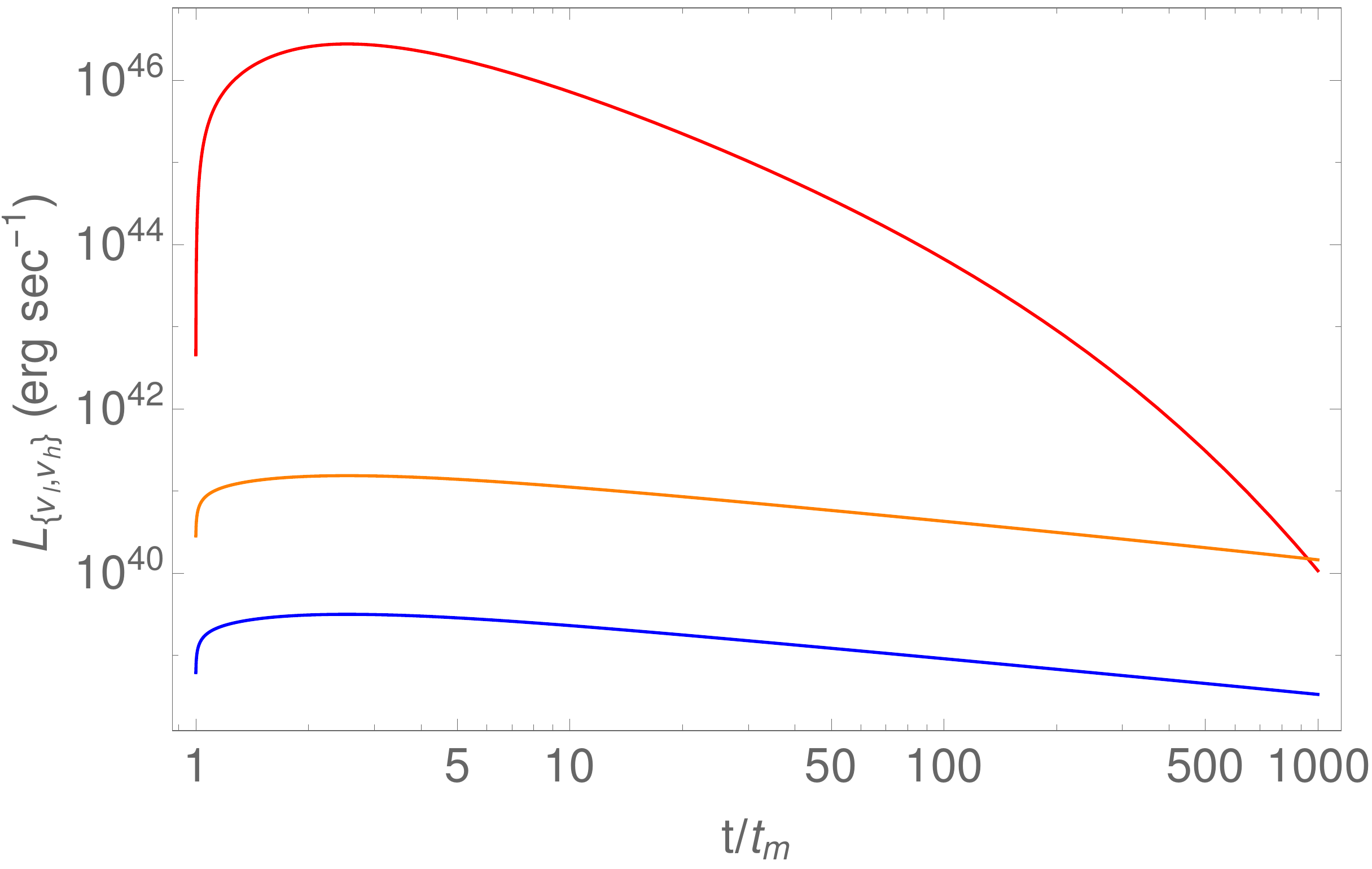}}
\caption{(a) The bolometric luminosity (eqn \ref{bolum}) fit  to  XMMSL1 J061927.1-655311 \citep{2014A&A...572A...1S} and the best fit parameters derived are $\bar{e}=10^{-4},~\ell=0.6,~M_6=6.5,~m=1$ and $j=0.3$ with a shift in time of 45 days in model to fit the observations. (b) The luminosity for the obtained fit parameters in the optical g band (5000-6000 $\AA$) (blue), UV (1660-2680 $\AA$) (orange) and X-rays (0.3-10 Kev) (red) with $t_m=0.33$ days.}
\label{lbfit}
\end{figure}

The rise time scale of bolometric luminosity is given by eqn (\ref{tpeak}) and in the initial stages is dominated by the X-ray luminosity, the fit of eqn (\ref{bolum}) to  XMMSL1 J061927.1-655311 \citep{2014A&A...572A...1S} is shown in Fig \ref{lbfit}.

The Eddington luminosity and the accretion rate are 

\begin{equation}
L_E=\frac{4\pi G M_{\bullet}c}{\kappa}=1.44 \times 10^{44}~M_6~{\rm erg~sec^{-1}};~~~~\dot{M}_E=\frac{L_E}{\eta c^2}=1.6\times 10^{24}\left(\frac{\eta}{0.1}\right) M_6 ~{\rm g~sec^{-1}},
\end{equation}

where $\kappa$ is the Thompson opacity and $\eta$ is the efficiency. From Fig \ref{lbfit}, we can see that the source is super-Eddington in initial stages and is sub-Eddington during the period of observations. The super-Eddington disks induces an outflowing wind and \citet{2009MNRAS.400.2070S} have proposed an adiabatic spherical model of the wind from a slim disk which is generalized by MM15 for the $\{E,~J\}$ phase space. The radius and the temperature of the photosphere of the wind are given by (MM15)

\begin{equation}
r_{\rm ph}= \frac{f_{\rm out}\dot{M}\kappa}{4\pi v_{w}};~~T_{\rm ph}=(4\pi)^{\frac{5}{12}}\left(\frac{1}{2a}\right)^{\frac{1}{4}}\kappa^{-\frac{2}{3}}f_{\rm out}^{-\frac{5}{12}}f_{v}^{\frac{11}{12}}\dot{M}^{-\frac{5}{12}}r_{c}^{-\frac{7}{24}}(GM_{\bullet})^{\frac{11}{24}}
\end{equation}

where $v_w=f_{v}\sqrt{G M_{\bullet}/r_c}$ is the velocity of the outflowing wind with $f_{v}$ taken to be unity and the fraction of mass outflow $f_{out}=\dot{M}_w/\dot{M}$ where $\dot{M}_w$ is the mass outflow rate, is given by (MM15)
 
\begin{equation}
f_{\rm out}=\frac{2}{\pi} \arctan \left[\frac{1}{4.5}\left(\frac{\dot{M}}{\dot{M}_E}-1\right)\right].
\end{equation}

From the slim disk model with $\alpha$ viscosity, the temperature of the disk is given by (MM15)

\begin{equation}
\sigma_{\rm SB} T_{e}^4=\frac{3}{8}\frac{GM_{\bullet}\dot{M}}{r^3}\left(1-\sqrt{\frac{r_{ISCO}}{r}}\right)\left[\frac{1}{2}+\left \{\frac{1}{4}+\frac{3}{2}\left(1-\sqrt{\frac{r_{ISCO}}{r}}\right) \left (\frac{\dot{M}}{\eta \dot{M}_{E}}\right)^2 \left(\frac{r}{R_s}\right)^{-2}\right \}^{\frac{1}{2}}\right]^{-1}.
\end{equation} 

The bolometric luminosity of the outflowing wind is $L_b^w=4\pi r_{ph}^2 \sigma_{SB} T_{ph}^4$, and of the disk is \\ $\displaystyle{L_b^d=\int_{r_{in}}^{r_{out}}\sigma_{SB}T_e^4 2\pi r \, {\rm dr}}$. Thus the total bolometric luminosity is $L_b=L_b^w+L_b^d$. The luminosity in the given spectral band using eqn (\ref{speclum}) is $L_{\{\nu_{l},~\nu_h\}}=L_{\{\nu_{l},~\nu_h\}}^d(T_e)+4\pi r_{ph}^2 \int_{\nu_{l}}^{\nu_{h}} B(T_{ph}) \, {\rm d\nu}$, where $B(T_{ph})=(2h/c^2) (\nu^3/({\rm Exp}[h\nu/(k_B T_{ph})]-1))$.

\subsection{A time dependent model}

The accretion disk models we have considered previously are steady accretion disks with time varying accretion rate taken to be mass fallback rate. 
The infall debris circularizes to form an accretion disk which can be a sub-Eddington or super-Eddington disk. The hydrodynamic equations of disk in the limit of $v_r \ll v_{\phi}$, where $v_r$ is the radial velocity and $v_{\phi}$ is the azimuthal velocity are given by 

\begin{subequations}
\begin{align}
\frac{\partial}{\partial t} \Sigma_d &=-\frac{1}{r}\frac{\partial}{\partial r}(r v_r \Sigma_d)-\dot{\Sigma}_w+\dot{\Sigma}_f \\
\omega^2 &=\frac{1}{r}\frac{\partial}{\partial r}\Phi(r) \\
v_r \Sigma_d\frac{\partial}{\partial r}(r^2 \omega)&+\dot{\Sigma}_w r^2 \omega(r)=-\frac{1}{r}\frac{\partial}{\partial r}(r^2 \Pi_{r\phi})+\dot{j}_f
\end{align}
\label{supdisk1}
\end{subequations} 

where $\dot{\Sigma}_f$ and $\dot{j}_f$, are the mass and angular momentum per unit area added to the disk by the fallback debris. \citet{MM17sub} (hereafter MM17) have constructed a self similar model of the time dependent accretion disk with fallback from disrupted debris and viscosity prescription $\Pi_{r\phi}=-K \Sigma_d^b r^d$ and derived $b$ and $d$ for an assumed pressure and density structure of disk. The self similar structure of the disk is taken to be $\Sigma_d=\Sigma_0 (t/t_0)^{\beta} g(\xi)$, $\xi=(r/r_0)(t/t_0)^{\alpha}$ and a power law solution $g(\xi)=A\xi^p$ is considered, where $\Sigma_0,~r_0,~t_0,~\beta,~\alpha,~A$ and $p$ are constants derived from self similar solutions. We have considered five models which includes the $\alpha$ viscosity sub-Eddington, the radiative viscosity sub-Eddington, the $\alpha$ viscosity super-Eddington, radiative viscosity super-Eddington and gravitational instability. The gravitational instability disk in a self gravitating disk is dominant when the surface density is high and in case of TDE disks we found that its luminosity is smaller than the typical observed luminosity and thus we exclude this possibility. The sub-Eddington radiative disk has strong radiative pressure which results in an extended disk and the thin disk approximation is no longer valid, but an extended disk structure for super-Eddington disk holds instead. We further found that the sub-Eddington radiative disk and the super-Eddington $\alpha$ disks do not have consistent solutions and thus we are left with two types of accretion disks which are $\alpha$ viscosity sub-Eddington (model A) and radiative viscosity super-Eddington with wind (model B); see MM17.  

Using a mass conservation equation $\dot{M}_d=\dot{M}_{fb}-\dot{M}_a-\dot{M}_w$, where $M_d$ is the disk mass, $\dot{M}_a$ the accretion rate onto the black hole and $\dot{M}_w$ the mass outflow rate leaving the disk, we have found that the outer radius increases in both models A and B. The bolometric luminosity in model A is $L_b \propto t^{-1.4}$ (MM17). 

For a super-Eddington disk, MM17 considered an extended atmosphere dominated by the radiation pressure and using the vertical momentum equation with an assumption that the atmosphere is in hydrostatic equilibrium upto a photospheric height $z_{ph}$, they have obtained the density structure $\rho(r,~z)$, and temperature profile $T_h(r,~z)$. The photospheric height is obtained using the Eddington approximation and the Eddington temperature is given by $T_h(r,~z_{ph})$ and assuming a blackbody emission, we found that the Eddington luminosity $L_E$ increases with time. At $z_{ph}$, the wind is launched and the outflowing density is given by $\dot{\Sigma}_w \propto r^{-\frac{7}{4}} \sqrt{\frac{T_{ph}}{T_E}-1+c_2} \propto r^{-7/4}(t/t_0)^{\delta}$ where $c_2$ is a constant, $\delta=-5/3$ and $T_{ph}$ is the temperature of the photosphere. The bolometric luminosity of the disk is $L_b^d \propto t^{-5/3}$ at late times and that of the wind is $L_b^w=\left(W_1^2 \left(\frac{t}{t_0}\right)^{2\delta}+1-c_2\right)^4 L_E$ where $W_1$ is a constant obtained from the disk solution.

\begin{figure}[ht]
\begin{center}
\scalebox{0.73}
{
\tikzstyle{decision} = [diamond, draw, text width=6em, text badly centered, node distance=5cm, inner sep=0pt]
\tikzstyle{block} = [rectangle, draw, text width=10em, text centered, rounded corners, minimum height=2em]
\tikzstyle{cloud} = [draw, ellipse,node distance=3cm, minimum height=2em]
\tikzstyle{container}=[draw, rectangle, inner sep=0.3cm, minimum height=3cm]
\def\bottom#1#2{\hbox{\vbox to #1{\vfill\hbox{#2}}}}
\tikzset{
  mybackground/.style={execute at end picture={
      \begin{scope}[on background layer]
        \node[] at (current bounding box.north){\bottom{1cm} #1};
        \end{scope}
    }},
}
\begin{tikzpicture}[mybackground={\Large Super-Eddington}]
\node [block] (phase1) {Sub-Eddington \\ Model A};
\node [block, text width=6em, right of=phase1,node distance=5.0cm] (ph1) {Without wind \\ Model T};
\node [block, text width=6em, right of=ph1,node distance=3.6cm] (ph2) {With wind \\ Model B};
\node [block, text width=6em, right of=ph2,node distance=3.5cm] (ph3) {Without Wind \\ Model T };
\node [block, right of=ph3,node distance=5.0cm] (phase3) {Sub-Eddington \\ Model A };
\node [container, fit=(ph1)(ph2)(ph3)](container){};
\draw [->, thick] (phase1) -- (ph1);
\draw [->, thick] (ph1) -- (ph2);
\draw [->, thick] (ph2) -- (ph3);
\draw [->, thick] (ph3) -- (phase3);
\end{tikzpicture}
}
\end{center}
\caption{ The evolutionary track of super-Eddington TDEs. The sub-Eddington TDEs will have a single phase which can be fit by model A. The dynamics in the individual phases and the transition between them decide the light curve profiles. A detailed model for an occasional jet can piggyback this scheme. }
\label{mevt}
\end{figure}
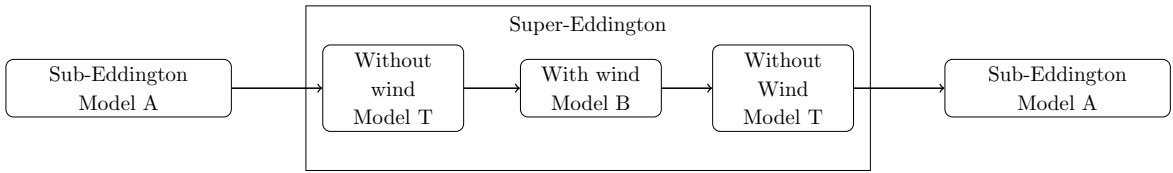

When the luminosity in model B goes below the Eddington luminosity, the disk should transit from model B to model A and the viscosity transits from radiative to $\alpha$ which has a lower heating rate. We found that there is a huge drop in the luminosity if the disk transits from model B to model A and this is because the viscosity transits from radiative to $\alpha$ viscosity. Thus the disk is taken to transit from model B to model T which is a super-Eddington disk without a wind and then to model A at late times. The evolutionary track of a super-Eddington disk is given in Fig \ref{mevt}. A model of jet formation which depends on $\dot{j}, \dot{M}, M_{BH}$ and $r_{in}$, is not included in the present scheme; since 
this event takes place in the inner magnetosphere, the model will not be impacted by a jet given by a detailed electrodynamical or magnetohydrodynamical model which can piggyback on this scenario. The applications of these models by
fits to observations of archival TDEs in X-ray and optical bands are discussed
in MM17 where both the Models A and B produce reasonable fits to the four diverse TDE sources considered. 

\section{Estimates of the expected rates of TDEs}

For a galactic core with stellar density $\rho(r) \propto r^{-\gamma}$ and the mass function $\xi(m){\rm d}m$ given in \citet{2001MNRAS.322..231K}, MM15 solved the steady state Fokker-Planck equation to obtain the capture rate $\dot{N}_t$ given by

\begin{equation}
\frac{{\rm d}^3 \dot{N}_{t}}{{\rm d} \bar{e} \hspace{1mm} {\rm d} \ell^2 \hspace{1mm}{\rm d} m}(\gamma,\bar{e},\ell,M_{\bullet},m)=4 \pi^2\hspace{2mm}s^{-1}_t\sigma^2 \hspace{2mm} \xi (m) \hspace{1mm} J^2_{lc}(\bar{e})\hspace{2mm} \mathcal{F}(\chi=1,\hspace{1mm}\ell),
\label{tdnth1}
\end{equation}

where $s_t=r_t/r_h,~J_{lc}^2(\bar{e})=2GM_{\bullet}r_t (1-\bar{e})$, $\mathcal{F}(\hspace{1mm}\ell)$ is the solution of Fokker-Planck equation, $\chi$ given by

\begin{equation}
\chi=\frac{1}{\left<D(\mathcal{E})\right>}\int_{r_{p}}^{r} \lim_{j\rightarrow 0}\frac{\left<(\Delta j_1)^2\right>}{2j_1 } \,\frac{{\rm d} r}{v_{r}}\hspace{5mm} {\rm and} \hspace{5mm} y=\frac{j_1}{\left<D(\mathcal{E})\right>},
\end{equation} 

where $\left<D(\mathcal{E})\right>$ is the orbital averaged angular momentum diffusion coefficient, $\left<(\Delta j_1)^2\right>$ is the diffusion coefficient, $v_r$ is the radial velocity of orbit, $j_1=J/J_c$ and $J_c$ the angular momentum of circular orbit. The net capture rate obtained by integrating eqn (\ref{tdnth1}) over phase space in the range $\bar{e}_h<\bar{e}<1,~0<\ell<1$ and $0.8<m<150$, is found to scale as $\dot{N}_t \propto M_6^{-0.3}$ and it is also found that $\dot{N}_t \propto \gamma^{2.1}$. 

MM15 used the steady accretion model given in \S \ref{lum} to simulate the light curve profiles in various spectral bands and the flux $f=L/(4\pi d_L(z)^2$ is compared with the sensitivity of the detector $f_l$ to obtain the duration of the flare detection $\delta t$, where the luminosity distance is given by 

\begin{equation}
{\rm d}_{L}(z)=(1+z) \frac{c}{H_o}  \int_0^z  \frac{1}{((1+z')^{3} \Omega_m+ \Omega_{\Lambda})^{0.5}} \,{\rm d} z',
\end{equation}

and $H_o,~\Omega_m$ and $\Omega_{\Lambda}$ are the cosmological parameters taken from the \citet{2014A&A...571A..29P}. If $\displaystyle{t_{\rm cad} \hspace{1mm} {\rm and} \hspace{1mm} t_{\rm int}}$ are the cadence and integration time of the detector, then the probability of detection of event is given by $P={\rm Min}\left[1, \delta t/(t_{cad}+t_{int})\right]$. The BH mass function of quiescent galaxies is (\citealt{2007ApJ...654..731H}; MM15)

\begin{equation}
\frac{{\rm d} \psi}{{\rm d} M_{\bullet}}=(1-\delta(z)) \frac{{\rm d} \psi}{{\rm d} \log L} \left|\frac{{\rm d} \log L}{{\rm d} M_{\bullet}}\right|;~~~\frac{{\rm d} \psi}{{\rm d} \log L}= \frac{\psi_{\ast}}{(\frac{L}{L_{\ast}})^{\gamma_1}+(\frac{L}{L_{\ast}})^{\gamma_2}},
\end{equation}

The comoving volume is

\begin{equation}
{\rm d} V_c=\omega d_H^3 \frac{I^2(z)}{W(z)} {\rm d} z;~~I(z)= \int_0^z  \frac{1}{((1+z')^{3} \Omega_m+ \Omega_{\Lambda})^{0.5}} \,{\rm d} z'
\label{dvol}
\end{equation}
\\
where $\omega=4\pi f_{s},\hspace{1mm} d_H=\displaystyle{c/H_o},\hspace{1mm} W(z)=((1+z)^{3} \Omega_m+ \Omega_{\Lambda})^{0.5}$ and $f_{s}$ is the fraction of sky observed. The net detectable rate of TDEs is given by

\begin{equation}
\displaystyle \dot{N}_{D}= \int^{100}_{1} {\rm d} M_6 \int^{150}_{0.8} {\rm d} m \int^{1}_{\bar{e}_{h}} {\rm d} \bar{e} \int^{1}_{0} {\rm d} \ell \int^{z_{s}}_{0} {\rm d} z  \hspace{2mm}\frac{1}{1+z}  \frac{{\rm d}^5\dot{N}_{D}(\gamma,\bar{e},\ell,M_6,m,z)}{{\rm d} M_6\hspace{0.2mm}{\rm d} m\hspace{0.2mm}{\rm d} \bar{e}\hspace{0.2mm}{\rm d} \ell \hspace{0.2mm}{\rm d} z}.
\label{det}
\end{equation}  

where 

\begin{equation}
 \frac{{\rm d}^5\dot{N}_{D}(\gamma,\bar{e},\ell,M_{\bullet},m,z)}{{\rm d} M_{\bullet}\hspace{0.2mm}{\rm d} m \hspace{0.2mm}{\rm d} \bar{e}\hspace{0.2mm}{\rm d} \ell \hspace{0.2mm}{\rm d} z}=\omega d^3_H \left(\frac{{\rm d} \Psi}{{\rm d} M_{\bullet}}\right) \frac{{\rm d}^3 \dot{N}_{t}}{{\rm d} \bar{e} \hspace{0.2mm} {\rm d} \ell \hspace{0.2mm} {\rm d} m}(\gamma,\bar{e},\ell,M_{\bullet},m) \frac{I^2(z)}{W(z)} P(\bar{e},\ell,M_{\bullet},m,z).
\label{net}
\end{equation}

Calculations of $\dot{N}_d$ for the various LSST, Pan-STARRS, and eROSITA surveys are illustrated in Table \ref{table1}. These estimates are in rough agreement with estimates of \citet{2011ApJ...741...73V} who scaled up SDSS results to other surveys.  

\begin{table}
\small
\begin{center} 
\begin{tabular}{|c|c|c|c|c|c|c|}
\hline 
Surveys & Band & Sensitivity/flux & $f_{sky}$ & $t_{cad}$ & $t_{int}$ & $\dot{N}_D$ \\
 &&&& sec & sec & ${\rm yr^{-1}}$\\
\hline
LSST & optical & 24.5 AB (g band) & 0.5 & 2.6 $\times 10^5$ & 10 & 5003 \\
&&&&&& \\
Pan-STARRS 3$\pi$ & optical & 24.0 AB (g band) & 0.75 & 6.05 $\times 10^5$ & 30 & 6337 \\
&&&&&& \\
Pan-STARRS MDS & optical & 24.8 AB (g band) & 0.0012 & 3.46 $\times 10^5$ & 30 & 12.3 \\
&&&&&& \\
eROSITA & X-ray & 2.4 $\times 10^{-14}$ (${\rm erg~sec^{-1}} $) & 1 & 1.58 $\times 10^7$ & 1.6 $\times 10^3$ & 679.5 \\
\hline 
\end{tabular} 
\end{center} 
\caption{Our estimated rates for TDE surveys (MM15). The medium deep survey (MDS) observes a small portion of sky with high sensitivity compared to the 3$\pi$ survey.
\label{table1}} 
\end{table} 

\section{TDE detections by observations}

The early discoveries of TDEs were made by the ROSAT surveys in 1990s in the form of X-ray outbursts from quiescent galaxies \citep{1996A&A...309L..35B,1999A&A...343..775K}. The TDEs in X-rays were also observed by XMM-Newton \citep{2007A&A...462L..49E,2014A&A...572A...1S} and Chandra X-ray observatory \citep{2010ApJ...722.1035M}. Apart from X-rays, the TDEs are also observed in the optical bands \citep{2008ApJ...676..944G,2014MNRAS.445.3263H} and in UV \citep{2008ApJ...676..944G}. The TDE candidates nowadays are followed up by multiwavelength observations and their high resolution spectra provide a deeper understanding of the accretion disks \citep{2014MNRAS.445.3263H,2017MNRAS.465L.114W}.  These observations were fit with the standard $t^{-5/3}$ model to deduce the black hole masses. The discovery of jetted TDEs \citep{2011Natur.476..421B} along with their radio counterpart \citep{2012ApJ...748...36B} have opened a new window in the field of TDE dynamics.

The TDEs detected by All Sky Survey missions are followed up by observations in X-ray, UV and optical bands using various ground based and space detectors. The emission from TDE disks are initially in the X-ray bands due to high disk temperature which lasts over a period of days to months then followed by peak emission in UV bands which lasts over a period of a few months and the optical emission which lasts over a period of year. Some TDEs are found to have radio counterparts which sustain over a time scale of few years and is associated with outflowing jets whose formation and evolution mechanisms are unclear. 

Once the TDE is detected and observed, the Difference Image Analysis (DIA) is performed in which the source image is subtracted from the reference image (observation prior to detection of transient) by modeling the alignment difference, the point spread function (PSF), exposure time, atmospheric extinction and sky background between them \citep{2014MNRAS.445.3263H,2017MNRAS.465L.114W}. All non varying sources are subtracted out leaving only the sources which varied over time. The source offset is measured by getting the position of the source from DIA and the centroid of galaxies based on the stacks of previous observations. The emission is considered to be nuclear if the source lies within a few arcseconds from the center. The photometric analysis is then performed on DIA source image to obtain the light curve profiles. 

Observations of the host galaxy prior to transient flare is necessary to decide the nature of galaxy's core. If there is no prior X-ray detection, then the galaxy core hosts a weak AGN or is quiescent. The spectra of weak AGN possess H$\alpha$, H$\beta$ weak narrow emission lines. The star formation is weak if the line ratio log[O{\sc iii}/H$\beta$] and log[N{\sc ii}/H$\alpha$] are small. The host galaxy spectrum is subtracted from the source spectrum to obtain the spectrum of the transient, i.e. TDE. The spectrum is dominated by H$\alpha$, H$\beta$ and sometimes He broad emission lines for TDEs \citep{2017MNRAS.465L.114W}. The TDE candidates such as PS1-10jh \citep{2012Natur.485..217G} and ASASSN-15oi \citep{2016MNRAS.463.3813H} show broad He emission lines with weak to no H emission lines at initial stages which implies that the disrupted star is He rich and hence possibly an evolved main sequence star.

\section{TDE surveys}

The search for TDEs has increased in a decade with the various ASS missions and follow ups from space and ground based detectors. The transient is detected through ongoing missions such as Monitor of All-sky X-ray Image (MAXI), Astrosat SSM in X-rays and Zwicky Transient Factory (ZTF; iPTF), Optical Gravitational Lensing Experiment (OGLE), All Sky Automated Survey for Supernova (ASAS SN) and Panoramic Survey Telescope and Rapid Response System (Pan-STARRS) in the optical band. The future surveys such as Large Synoptic Survey Telescope (LSST) in optical and eROSITA in X-rays will further boost the detection rate of TDEs. MM15 through detailed modeling of stellar and accretion dynamics have obtained the detection rate of TDEs for LSST to be $\sim 5000~{\rm yr^{-1}}$, Pan-STARRS to be $\sim 6337~{\rm yr^{-1}}$ and eROSITA to be $\sim 679~{\rm yr^{-1}}$.

\begin{figure}[ht]
\begin{center}
\scalebox{0.9}
{
\tikzstyle{decision} = [diamond, draw,  
    text width=20em, text badly centered, node distance=5cm, inner sep=0pt]
\tikzstyle{block} = [rectangle, draw, 
    text width=27em,  rounded corners, minimum height=2em]
\tikzstyle{line} = [draw, -, line width=0.102em]
\tikzstyle{arr} = [draw, -latex', line width=0.102em]
\tikzstyle{cloud} = [draw, ellipse,node distance=3cm,
    minimum height=2em]

\begin{tikzpicture}[node distance = 1.4cm, auto]
\coordinate (A) at (-9.5,1);
\coordinate (B) at ( -9.5,-1.5);
\coordinate (C) at ( -9.5,-2.6);
\coordinate (D) at ( -9.5,-5.2);
\coordinate (E) at ( -9.5,-6.8);
\coordinate (F) at ( -9.5,-9.0);
\draw[line] (A) -- (B) node[midway] {$t = 0$};
\draw[line] (B) -- (C) node[midway] {$\sim$ Few Months};
\draw[line] (C) -- (D) node[midway] {$\sim$ Months to Year};
\draw[line] (D) -- (E) node[midway] {$\sim$ Year};
\draw[arr] (E) -- (F) node[midway] {$\sim$ Few Years};

\node [text width=3em] at ( -9.3,-9.3) {Time};
\node [block] (alert) {{\bf Alert:} X-ray: MAXI, eROSITA-SRG, \textcolor{blue}{ASTROSAT SSM} \\~~~~~~~~~ Optical: iPTF/zPTF, OGLE, ASAS-SN, LSST, \\~~~~~~~~~~~~~~~~~~~~~ Pan-STARRS, \textcolor{blue}{ILMT}};
\node [block, below of=alert, node distance=2cm] (xray) {{\bf X-ray:} XMM-EPIC, SWIFT XRT, \textcolor{blue}{ASTROSAT SXT},\\~~~~~~~~~~ Chandra X-ray  };
\node [block, below of=xray, node distance=2cm] (uv) {{\bf UV:} XMM-OM, SWIFT UVOT, \textcolor{blue}{ASTROSAT UVIT} };
\node [block, below of=uv, node distance=2cm] (optical) {{\bf Optical:} XMM-OT, SWIFT UVOT, Pan STARRS, \textcolor{blue}{ILMT}, \\ \textcolor{blue}{2m HCT}, \textcolor{blue}{Devasthal Optical Telescope, ASTROSAT UVIT} };
\node [block, below of=optical, node distance=2cm] (radio) {{\bf Radio:} VLBA, \textcolor{blue}{GMRT}};
\end{tikzpicture}
}
\end{center}
\caption{The detection and follow up missions for TDEs are shown along with the typical timescale of observations in various spectral bands starting from the time of the alert. The blue color highlights Indian observatories. }
\label{tdes}
\end{figure}
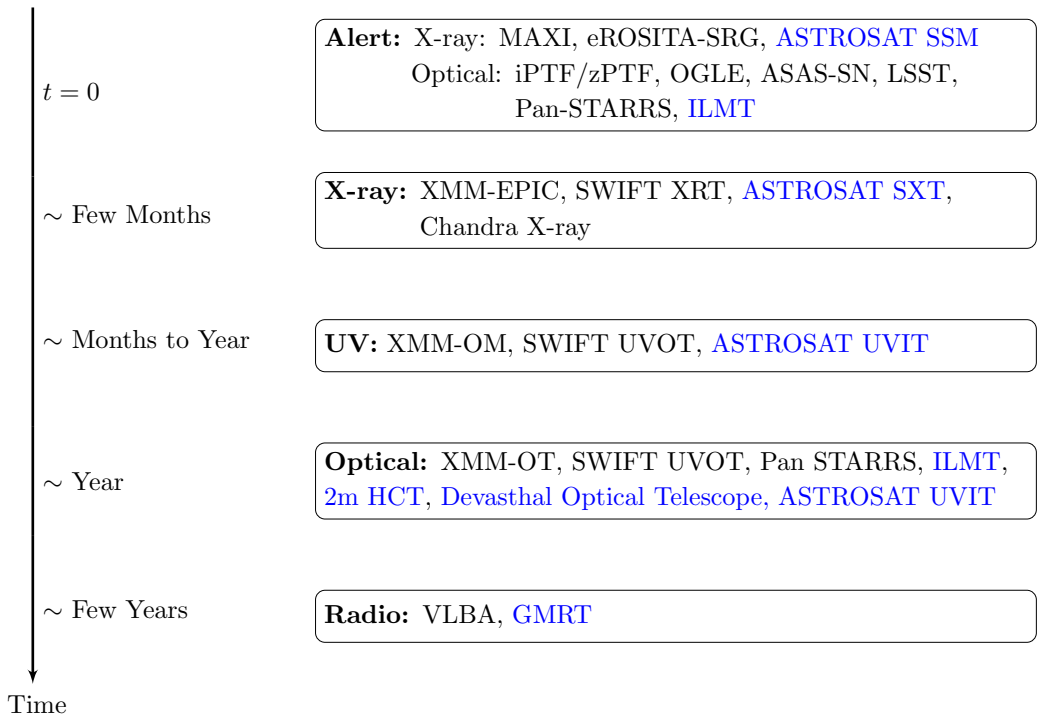

The follow up of detected TDEs is done in X-rays by XMM-Newton, Swift XRT and Chandra X-ray telescope, in UV bands by XMM-Newton Optical Monitors and Swift UVOT and in optical by XMM-Newton Optical Monitors, Swift UVOT and Pan-STARRS as shown in Fig \ref{tdes}.

India's multiwavelength space mission ASTROSAT has payloads such as SSM for monitoring the sky in nearly every six hours with a sensitivity of $\sim$30 mcrab, besides UVIT to observe in far ultraviolet (FUV) and near ultraviolet (NUV), and soft X-ray telescope (SXT) to observe in X-ray 0.2-10 kev bands. The 4m International Liquid Mirror Telescope (ILMT) in Devasthal, India is entirely dedicated to photometry and astrometry direct imaging surveys. ILMT points toward the zenith and scans the strip of sky every night with limiting magnitude of 22 in i band and sub-arcsecond resolution. With a cadence of one night and higher sensitivity in single integration, ILMT provides a potential mission for TDE search. The 3.6 m Devasthal Optical Telescope (DOT) observes in optical and near infrared bands with a high sensitivity. The 4k $\times$ 4k CCD imager (5 -10 arcmin) covers UVBRI bands with a limiting magnitude of 24 in the R band. The 2 m Himalayan Chandra Telescope (HCT) performs imaging in Bessell UBVRI bands with a limiting magnitue of 22.2 in R band and spectroscopy with a resolution of 300 and limiting magnitude of 18.5 in V band using HFOSC. The Hanle Echelle Spectrograph (HESP) in HCT covers the entire optical wavelength with a spectral resolution of 30000 and 60000.

\section{Conclusions: Theory, TDE search and follow up}

Tidal disruption events are important for black hole demographics such as deriving distributions
in mass and redshift, besides being a key laboratory for accretion and jet physics and stellar dynamics.
We can also make inferences about the properties of disrupted stars. On the theoretical side, we are developing models for relativistic loss cone theory, resonant interactions in stellar diffusion, and binary black holes.
Given that a high detectable event rate will become possible soon with eROSITA in X-rays and iPTF/ZTF, ASAS SN
and Pan-STARRS in the optical, there are several opportunities available currently and in the future. In the Indian
context, it is possible that ILMT can be used for picking up these events. Once a trigger is  received, there could
be follows up in X-rays (SXT), UV (UVIT), Optical (DOT, HCT) and in Radio (GMRT) at the appropriate times given in Fig. \ref{tdes}. With the DOT we can probe  TDEs longer and also study fainter sources. A key input in determining the
timing of the observations are the parameters extracted  using formulae presented here for rise time,
peak luminosity and decay time. There is a need to plan multiwavelength ToO campaign nationally with support
of ASTROSAT, HTAC, DOT, and GMRT TACs. As it is important in these studies to reduce the time gap between the alert and the observations, it would be ideal and imperative if the time allocation is made quickly when an opportunity presents itself when a TDE science team  makes a proposal for coordinated search and follow up.

%
\section*{Acknowledgements}

We thank Profs. G. C. Anupama and P. Sreekumar for discussions and their encouragement during this work.

\bibliography{reference}

\end{document}